\RequirePackage{fontspec}

\PassOptionsToPackage{unicode}{hyperref}
\PassOptionsToPackage{hyphens}{url}
\documentclass[
]{article}
\usepackage{xcolor}
\usepackage{amsmath,amssymb}
\usepackage{iftex}
\usepackage{newunicodechar}
\newunicodechar{≥}{\ensuremath{\geq}}
\newunicodechar{≤}{\ensuremath{\leq}}
\newunicodechar{𝒟}{\ensuremath{\mathcal{D}}}
\newunicodechar{♭}{\ensuremath{\flat}}
\newunicodechar{♯}{\ensuremath{\sharp}}
\newunicodechar{α}{\ensuremath{\alpha}}
\newunicodechar{β}{\ensuremath{\beta}}
\newunicodechar{γ}{\ensuremath{\gamma}}
\newunicodechar{δ}{\ensuremath{\delta}}
\newunicodechar{ε}{\ensuremath{\epsilon}}
\newunicodechar{ρ}{\ensuremath{\rho}}
\newunicodechar{σ}{\ensuremath{\sigma}}
\newunicodechar{Δ}{\ensuremath{\Delta}}
\newunicodechar{→}{\ensuremath{\rightarrow}}
\newunicodechar{←}{\ensuremath{\leftarrow}}
\newunicodechar{−}{\ensuremath{-}}
\newunicodechar{×}{\ensuremath{\times}}
\newunicodechar{∈}{\ensuremath{\in}}
\newunicodechar{∥}{\ensuremath{\parallel}}
\newunicodechar{∑}{\ensuremath{\sum}}
\newunicodechar{∞}{\ensuremath{\infty}}
\newunicodechar{≈}{\ensuremath{\approx}}
\newunicodechar{≠}{\ensuremath{\neq}}
\newunicodechar{·}{\ensuremath{\cdot}}
\newunicodechar{✓}{$\checkmark$}
\ifPDFTeX
  \usepackage[T1]{fontenc}
  \usepackage[utf8]{inputenc}
  \usepackage{textcomp} 
\else 
  \usepackage{unicode-math} 
  \defaultfontfeatures{Scale=MatchLowercase}
  \defaultfontfeatures[\rmfamily]{Ligatures=TeX,Scale=1}
\fi
\usepackage{lmodern}
\ifPDFTeX\else
\fi
\IfFileExists{upquote.sty}{\usepackage{upquote}}{}
\IfFileExists{microtype.sty}{
  \usepackage[]{microtype}
  \UseMicrotypeSet[protrusion]{basicmath} 
}{}
\makeatletter
\@ifundefined{KOMAClassName}{
  \IfFileExists{parskip.sty}{%
    \usepackage{parskip}
  }{
    \setlength{\parindent}{0pt}
    \setlength{\parskip}{6pt plus 2pt minus 1pt}}
}{
  \KOMAoptions{parskip=half}}
\makeatother
\usepackage{longtable,booktabs,array}
\usepackage{calc} 
\usepackage{etoolbox}
\makeatletter
\patchcmd\longtable{\par}{\if@noskipsec\mbox{}\fi\par}{}{}
\makeatother
\IfFileExists{footnotehyper.sty}{\usepackage{footnotehyper}}{\usepackage{footnote}}
\makesavenoteenv{longtable}
\usepackage{graphicx}
\makeatletter
\newsavebox\pandoc@box
\newcommand*\pandocbounded[1]{
  \sbox\pandoc@box{#1}%
  \Gscale@div\@tempa{\textheight}{\dimexpr\ht\pandoc@box+\dp\pandoc@box\relax}%
  \Gscale@div\@tempb{\linewidth}{\wd\pandoc@box}%
  \ifdim\@tempb\p@<\@tempa\p@\let\@tempa\@tempb\fi
  \ifdim\@tempa\p@<\p@\scalebox{\@tempa}{\usebox\pandoc@box}%
  \else\usebox{\pandoc@box}%
  \fi%
}
\def\fps@figure{htbp}
\makeatother
\NewDocumentCommand\citeproctext{}{}
\NewDocumentCommand\citeproc{mm}{%
  \begingroup\def\citeproctext{#2}\cite{#1}\endgroup}
\makeatletter
 \let\@cite@ofmt\@firstofone
 \def\@biblabel#1{}
 \def\@cite#1#2{{#1\if@tempswa , #2\fi}}
\makeatother
\newlength{\cslhangindent}
\newlength{\csllabelwidth}
\newenvironment{CSLReferences}[2] 
 {\begin{list}{}{%
  \setlength{\itemindent}{0pt}
  \setlength{\leftmargin}{0pt}
  \setlength{\parsep}{0pt}
  \ifodd #1
   \setlength{\leftmargin}{\cslhangindent}
   \setlength{\itemindent}{-1\cslhangindent}
  \fi
  \setlength{\itemsep}{#2\baselineskip}}}
 {\end{list}}
\usepackage{calc}

\providecommand{\tightlist}{%
  \setlength{\itemsep}{0pt}\setlength{\parskip}{0pt}}
\usepackage{hyperref} 
\usepackage{bookmark}
\IfFileExists{xurl.sty}{\usepackage{xurl}}{} 
\hypersetup{
  pdftitle={vega-mir: An information-theoretic Python toolkit for symbolic music{,} with applications to harmonic graphs and rubato spectra},
  pdfauthor={Fred Jalbert-Desforges},
  hidelinks,
  pdfcreator={LaTeX via pandoc}}

\title{vega-mir: An information-theoretic Python toolkit for symbolic
music, with applications to harmonic graphs and rubato spectra}
\author{Fred Jalbert-Desforges}
\date{May 2026}

\begin{document}
\maketitle

\section{vega-mir: An information-theoretic Python toolkit for symbolic
music, with applications to harmonic graphs and rubato
spectra}\label{vega-mir-an-information-theoretic-python-toolkit-for-symbolic-music-with-applications-to-harmonic-graphs-and-rubato-spectra}

\textbf{Fred Jalbert-Desforges} \emph{Independent Researcher, Cygnus
Analysis · Montréal, Québec}
\href{mailto:fred@irsac.org}{\nolinkurl{fred@irsac.org}} ·
\href{https://fredjalbertdesforges.com}{fredjalbertdesforges.com} ·
ORCID \href{https://orcid.org/0009-0002-4357-6942}{0009-0002-4357-6942}

\emph{May 2026}

\textbf{Keywords}: music information retrieval, information theory,
symbolic music, Python, network analysis, rubato spectral analysis,
Higuchi fractal dimension

\begin{center}\rule{0.5\linewidth}{0.5pt}\end{center}

\subsection{Abstract}\label{abstract}

We present \texttt{vega-mir}, an open-source Python library that bundles
nine information-theoretic and statistical metrics for the analysis of
symbolic music corpora behind a small, tested, citable API, and
demonstrates two of them at corpus scale in case studies not addressed
by the upstream Cygnus paper. Of the nine metrics, three (Shannon
entropy, Kullback-Leibler divergence, Zipfian fits) were deployed in the
companion Cygnus arXiv preprint; two (network analysis on
chord-transition graphs and spectral analysis of rubato curves) are
deployed in full case studies here; the four remaining
(multi-dimensional Gini, chi-squared stationarity, Higuchi fractal
dimension, interval distribution) are validated against analytic anchors
and exercised as sanity checks on a bundled \(8\)-composer dataset. The
two case studies yield two main observations. First, on the fourteen
MAESTRO composers with \(N \geq 10\) pieces, the PageRank value of the
gravity-centre node correlates with the marginal Kullback-Leibler
distance at \(\rho = 0.61\) (Spearman, composer-level jackknife
\(N = 14\)); the categorical gravity-centre identity takes five distinct
values across the corpus but is not itself correlated with marginal KL
(\(\rho = 0.13\), \(p = 0.21\)). Second, on the \(247\)-piece Bach
multi-master corpus (Schiff, Gould, Richter), Gould holds the
\emph{highest} periodicity ratio of the three performers, not the
lowest, inverting the cliché that low scalar rubato reads as
``metronomic'': Gould's rubato is small in amplitude but structured in
time, with a median dominant period of \(66\) beats against Schiff's
\(102\) and Richter's \(104\).

\begin{center}\rule{0.5\linewidth}{0.5pt}\end{center}

\section{1. Introduction}\label{introduction}

Information-theoretic descriptors have become standard tools for the
quantitative analysis of symbolic music: Shannon entropy on
distributions of scale degrees, Kullback-Leibler divergence between
composer corpora, Zipfian fits on rank-frequency profiles, fractal
dimension on note-density curves. Yet implementations of these
quantities remain fragmented across the literature. Each new study
typically stitches together calls to \texttt{scipy.stats.entropy},
hand-rolled Laplace smoothing, custom Kullback-Leibler formulas, and
ad-hoc NetworkX code, with no shared API, no shared test suite, and no
shared reference values against which results can be cross-validated.

This paper presents \texttt{vega-mir}, an open-source Python library
that bundles nine information-theoretic and statistical metrics for the
analysis of symbolic music corpora behind a small, tested, citable API,
and demonstrates two of them at corpus scale in case studies that the
upstream Cygnus paper
(\citeproc{ref-jalbert2026cygnus}{Jalbert-Desforges 2026}) did not
address. Of the nine metrics, three (Shannon entropy, asymmetric
Kullback-Leibler divergence, Zipfian fits) were deployed in
(\citeproc{ref-jalbert2026cygnus}{Jalbert-Desforges 2026}) and are
documented here as part of the catalogue; two (network analysis on
chord-transition graphs and spectral analysis of rubato curves) are
deployed in full case studies in §§4 and 5; the four remaining
(multi-dimensional Gini, chi-squared stationarity, Higuchi fractal
dimension, interval distribution) are validated against analytic anchors
in the test suite and exercised as sanity checks on the bundled
\(8\)-composer dataset in §3.8. Their empirical deployment at full
corpus scale is left to future work. The library targets researchers in
music information retrieval, computational musicology, and
complex-systems studies of music who need reproducible distributional
analyses without reimplementing the underlying mathematics for each
project.

\texttt{vega-mir} operates on \textbf{symbolic} input (sequences of
scale degrees, chord-transition graphs, tempo curves, distributions over
discrete alphabets) and complements the upstream Cygnus pipeline
(\citeproc{ref-jalbert2026cygnus}{Jalbert-Desforges 2026}), which
derives such symbolic representations from raw audio recordings of piano
performances and reports a certified note-level transcription accuracy
of \(F_1 = 0.9791\) on the MAESTRO v3.0.0 test set. The companion paper
(\citeproc{ref-jalbert2026cygnus}{Jalbert-Desforges 2026}) establishes
the audio-to-analysis chain end-to-end on 1,238 MAESTRO pieces and 111
neoclassical recordings, and demonstrates three of the nine metrics
(Shannon entropy, asymmetric Kullback-Leibler divergence, Zipfian fits)
at corpus scale. The present paper makes three contributions:

\begin{enumerate}
\def\labelenumi{\arabic{enumi}.}
\item
  \textbf{Library presentation} (§2): the design, API, and validation
  strategy of \texttt{vega-mir} v0.0.1, archived on Zenodo with a
  citable DOI
  (\href{https://doi.org/10.5281/zenodo.19711033}{\texttt{10.5281/zenodo.19711033}})
  and built within a cross-platform continuous-integration matrix.
\item
  \textbf{Metric catalogue} (§3): the nine metrics, each with
  definition, theoretical anchor, default parameters matching the Cygnus
  methodology, and indication of which metrics were covered by the
  upstream Cygnus paper
  (\citeproc{ref-jalbert2026cygnus}{Jalbert-Desforges 2026}) versus
  those introduced here for the first time.
\item
  \textbf{Two case studies} (§§4--5): applications of metrics not
  addressed in (\citeproc{ref-jalbert2026cygnus}{Jalbert-Desforges
  2026}). First (§4), a network analysis of chord-transition graphs on
  the fourteen MAESTRO composers with \(N \geq 10\) pieces yields two
  complementary observations on the relationship between marginal and
  transition-level harmonic structure. The PageRank value of the
  gravity-centre node (a continuous scalar) correlates with the marginal
  Kullback-Leibler distance at \(\rho = 0.61\) (Spearman, composer-level
  jackknife \(N = 14\)); the categorical gravity-centre identity takes
  five distinct values across the corpus but is not itself correlated
  with marginal KL (\(\rho = 0.13\), \(p = 0.21\)). Second (§5), a
  spectral analysis of rubato curves across the \(247\)-piece Bach
  multi-master corpus (Schiff \(82\), Gould \(112\), Richter \(53\))
  recovers a temporal-colour axis hidden by the scalar mean: Gould has
  the \emph{highest} periodicity ratio of the three pianists, not the
  lowest, inverting the cliché that low scalar rubato reads as
  ``metronomic''.
\end{enumerate}

The intent is not to introduce new mathematics: every metric in
\texttt{vega-mir} has decades of precedent in information theory,
complex systems, and signal processing. The contribution is
consolidation, a single tested API with consistent defaults and a shared
set of reference values, lowering the activation cost for distributional
analyses of symbolic music corpora and enabling direct comparison of
results across studies.

\section{2. Library design}\label{library-design}

\texttt{vega-mir} targets Python 3.10 and above, with mandatory
dependencies on \texttt{numpy} (\citeproc{ref-harris2020numpy}{{Harris
et al.} 2020}), \texttt{scipy}
(\citeproc{ref-virtanen2020scipy}{{Virtanen et al.} 2020}),
\texttt{networkx} (\citeproc{ref-hagberg2008networkx}{Hagberg et al.
2008}), \texttt{mido}, and \texttt{pretty\_midi}. The package uses the
modern \texttt{src/} layout with \texttt{hatchling} as the build
backend. Source code, tests, and notebooks are released under the MIT
license at
\href{https://github.com/fredericjalbertdesforges/vega-mir}{github.com/fredericjalbertdesforges/vega-mir}.

\subsection{2.1 API surface}\label{api-surface}

The public API follows a uniform three-layer pattern per metric:

\begin{itemize}
\tightlist
\item
  a \textbf{primitive function} that operates on a normalised
  probability vector or a 1-D time series;
\item
  a \textbf{counts-based convenience} that applies Laplace smoothing on
  a fixed alphabet;
\item
  a \textbf{sequence-based convenience} for the most common chord or
  scale-degree input, with built-in consecutive-duplicate collapsing.
\end{itemize}

Each computation that returns more than a single scalar is wrapped in a
\texttt{NamedTuple} for structured, immutable, type-checkable access:
\texttt{ZipfFit}, \texttt{IntervalAnalysis}, \texttt{NetworkAnalysis},
\texttt{RubatoAnalysis}, \texttt{StationarityResult},
\texttt{FractalDimension}, \texttt{DominantPeriod}. Type hints are
present throughout, and \texttt{mypy\ -\/-strict} runs on every push.
Defaults match the Cygnus methodology
(\citeproc{ref-jalbert2026cygnus}{Jalbert-Desforges 2026}):
Jeffreys-Laplace smoothing with \(\alpha = 0.5\), log base 2,
consecutive-duplicate collapsing for transition-sequence analyses.
Degenerate inputs (too-short sequences, a single distinct symbol,
sub-threshold sample sizes) raise \texttt{ValueError} with explicit
messages rather than returning silent zeros, and the tests cover those
boundaries explicitly.

\subsection{2.2 Validation strategy}\label{validation-strategy}

Each metric is unit-tested against two independent ground truths. First,
\textbf{analytic anchors on canonical inputs}: a uniform distribution on
\(N\) symbols recovers \(H = \log_2 N\); a perfect Zipfian sequence
generated as \(P_i \propto 1/i\) recovers \(\alpha = 1\) with
\(R^2 = 1\); the Higuchi fractal dimension on white noise approaches
\(D = 2\); the Kullback-Leibler divergence between identical
distributions returns \(0\); and so on for all nine metrics. Second,
\textbf{exact parity against the upstream Cygnus reference
implementation} (\citeproc{ref-jalbert2026cygnus}{Jalbert-Desforges
2026}), for both raw computation and post-Laplace smoothing.

The 181-test suite runs in under one second. A continuous-integration
matrix on GitHub Actions exercises the suite on \texttt{ubuntu-latest},
\texttt{macos-latest}, and \texttt{windows-latest} against Python 3.10
through 3.13 with \texttt{ruff}, \texttt{mypy\ -\/-strict}, and
\texttt{pytest}, all currently green. Releases are archived on Zenodo
with a citable DOI; the v0.0.1 release {[}DOI
\href{https://doi.org/10.5281/zenodo.19711033}{\texttt{10.5281/zenodo.19711033}}{]}
is the version documented in this paper.

\subsection{2.3 Comparison with adjacent
libraries}\label{comparison-with-adjacent-libraries}

Several mature Python and Java libraries cover adjacent problems in
symbolic music research. \texttt{music21}
(\citeproc{ref-cuthbert2010music21}{Cuthbert and Ariza 2010}) is a
comprehensive computational-musicology framework with strong parsing,
harmonic analysis, and key-detection facilities, but its analytical
surface does not expose dedicated entropy, divergence, or power-law
fitting primitives. \texttt{partitura}
(\citeproc{ref-cancino2022partitura}{Cancino-Chacón et al. 2022})
specialises in lossless score parsing and performance alignment, leaving
downstream statistical analysis to the user. \texttt{jSymbolic}
(\citeproc{ref-mckay2018jsymbolic}{McKay et al. 2018}) is a
comprehensive Java-based feature extractor whose primary focus is broad
histogram-based descriptors, and which is not directly callable from a
Python pipeline. \texttt{musif} (\citeproc{ref-llorens2023musif}{Llorens
et al. 2023}) targets stylometric feature extraction with a
similarity-classification framing rather than an information-theoretic
one; entropy and divergence are not first-class outputs.
\texttt{vega-mir} complements these libraries by consolidating
information-theoretic computations behind a single tested API with
consistent defaults and a shared set of reference values, allowing
direct comparison of results across studies.

\section{3. The nine metrics}\label{the-nine-metrics}

This section catalogues the nine metrics implemented in
\texttt{vega-mir}, grouped by underlying mathematical domain. For each
metric we give the definition, the canonical reference, the entry-point
function, and a one-line theoretical anchor used in validation. Metrics
marked \textbf{(Cygnus)} were applied at corpus scale in the companion
paper (\citeproc{ref-jalbert2026cygnus}{Jalbert-Desforges 2026}); those
marked \textbf{(new)} are introduced here.

\subsection{3.1 Information theory}\label{information-theory}

\textbf{Shannon entropy} (Cygnus). For a discrete distribution \(P\) on
alphabet \(\mathcal{D}\) of size \(N\), the Shannon entropy is
\(H(P) = -\sum_i P(i) \log_2 P(i)\)
(\citeproc{ref-shannon1948mathematical}{Shannon 1948}). Entry point:
\texttt{shannon\_scale\_degrees}. Anchor: a uniform distribution on
\(N\) symbols returns \(\log_2 N\).

\textbf{Kullback-Leibler divergence} (Cygnus). Asymmetric divergence
between two distributions \(P\) and \(Q\) on the same alphabet,
\(D_{\mathrm{KL}}(P \| Q) = \sum_i P(i) \log_2 [P(i) / Q(i)]\), with
optional Jensen-Shannon symmetrisation
\(D_{\mathrm{JS}}(P, Q) = \frac{1}{2} D_{\mathrm{KL}}(P \| M) +
\frac{1}{2} D_{\mathrm{KL}}(Q \| M)\), \(M = \frac{1}{2}(P + Q)\). Entry
points: \texttt{kl\_divergence}, \texttt{js\_divergence},
\texttt{kl\_pairwise\_matrix}. Anchor: \(D_{\mathrm{KL}}(P \| P) = 0\);
bootstrap confidence intervals provided.

\subsection{3.2 Rank-frequency law}\label{rank-frequency-law}

\textbf{Zipf's law} (Cygnus). Power-law fit \(f_r \propto 1/r^\alpha\)
to a rank-frequency profile (\citeproc{ref-zipf1949human}{Zipf 1949}),
computed separately on marginal and joint (transition) distributions.
Entry point: \texttt{zipf\_fit}. Anchor: a perfectly Zipfian sequence
recovers \(\alpha = 1\) with \(R^2 = 1\); goodness-of-fit reported as
\(R^2\) in log-log space.

\subsection{3.3 Inequality}\label{inequality}

\textbf{Multi-dimensional Gini} (new). Gini coefficient applied to each
dimension of a multi-modal descriptor matrix (harmonic, dynamic,
rhythmic), yielding a per-dimension inequality score. Entry point:
\texttt{gini}. Anchor: a uniform distribution returns \(0\); a one-hot
distribution returns \(1 - 1/N\), approaching \(1\) for large \(N\).

\subsection{3.4 Networks}\label{networks}

\textbf{Network analysis on chord graphs} (new). Construction of the
directed weighted graph of chord-to-chord transitions and computation of
node-level (PageRank, in/out degree, clustering coefficient) and
graph-level (modularity-based community detection, density) descriptors
via NetworkX (\citeproc{ref-hagberg2008networkx}{Hagberg et al. 2008}).
Entry point: \texttt{network\_analysis}. Anchor: a uniform random graph
on \(N\) nodes with edge probability \(p\) returns clustering
coefficient \(\approx p\) and modularity vanishing as \(1/\sqrt{N}\).

\subsection{3.5 Time-series statistics}\label{time-series-statistics}

\textbf{Stationarity test} (new). Chi-squared test of independence of
the harmonic profile across consecutive segments of a piece, with
Cramer's V effect-size measure. Entry point:
\texttt{stationarity\_test}. Anchor: a stationary process returns
\(p > 0.05\) with \(V \to 0\); a piecewise-distinct process returns
\(p < 0.001\) with \(V \to 1\).

\textbf{Higuchi fractal dimension} (new). Fractal dimension of a 1-D
time series via curve length at multiple time scales
(\citeproc{ref-higuchi1988approach}{Higuchi 1988}). Entry point:
\texttt{higuchi\_fractal\_dimension}. Anchor: white noise approaches
\(D = 2\); a linear trend approaches \(D = 1\).

\textbf{Spectral rubato analysis} (new). Welch power spectral density on
a tempo curve, with classification of rubato into four categories
(steady, slow-oscillation, fast-oscillation, broadband noise) based on
dominant frequency and spectral entropy. Entry point:
\texttt{rubato\_spectral}. Anchor: a sinusoidal tempo modulation
recovers its modulation frequency as the dominant peak.

\subsection{3.6 Distribution shape}\label{distribution-shape}

\textbf{Interval distribution} (new). Statistical fit to melodic
intervals (Exponential or Laplace family) with goodness-of-fit reported
as log-likelihood and Kolmogorov-Smirnov distance. Entry point:
\texttt{interval\_analysis}. Anchor: synthetic intervals drawn from
known parameters recover those parameters within bootstrap confidence
intervals.

\subsection{3.7 Coverage summary}\label{coverage-summary}

{\def\LTcaptype{none} 
\begin{longtable}[]{@{}
  >{\raggedleft\arraybackslash}p{(\linewidth - 8\tabcolsep) * \real{0.2000}}
  >{\raggedright\arraybackslash}p{(\linewidth - 8\tabcolsep) * \real{0.1500}}
  >{\raggedright\arraybackslash}p{(\linewidth - 8\tabcolsep) * \real{0.1500}}
  >{\centering\arraybackslash}p{(\linewidth - 8\tabcolsep) * \real{0.2500}}
  >{\centering\arraybackslash}p{(\linewidth - 8\tabcolsep) * \real{0.2500}}@{}}
\toprule\noalign{}
\begin{minipage}[b]{\linewidth}\raggedleft
\#
\end{minipage} & \begin{minipage}[b]{\linewidth}\raggedright
Metric
\end{minipage} & \begin{minipage}[b]{\linewidth}\raggedright
Domain
\end{minipage} & \begin{minipage}[b]{\linewidth}\centering
In Cygnus paper
\end{minipage} & \begin{minipage}[b]{\linewidth}\centering
In this paper
\end{minipage} \\
\midrule\noalign{}
\endhead
\bottomrule\noalign{}
\endlastfoot
1 & Shannon entropy & Information theory & ✓ & catalogue \\
2 & Kullback-Leibler divergence & Information theory & ✓ & catalogue \\
3 & Zipfian fit (marginal + transitions) & Rank-frequency & ✓ &
catalogue \\
4 & Multi-dimensional Gini & Inequality & & catalogue \\
5 & Network analysis on chord graphs & Graph theory & & \textbf{§4 case
study} \\
6 & Stationarity test & Time series & & catalogue \\
7 & Higuchi fractal dimension & Complexity & & catalogue \\
8 & Spectral rubato analysis & Signal processing & & \textbf{§5 case
study} \\
9 & Interval distribution & Statistics & & catalogue \\
\end{longtable}
}

Three of the nine metrics (Shannon entropy, Kullback-Leibler divergence,
Zipfian fits) were established at corpus scale by the companion paper
(\citeproc{ref-jalbert2026cygnus}{Jalbert-Desforges 2026}). The
remaining six are introduced here, with two of them, network analysis on
chord graphs and spectral rubato analysis, deployed in full case studies
(§§4-5). The four others (Gini, stationarity, Higuchi, intervals) are
exercised as sanity checks on a bundled \(8\)-composer dataset in §3.8
and validated against analytic anchors in the test suite; their
empirical deployment at full corpus scale is left to future work.

\subsection{3.8 Sanity checks on bundled
data}\label{sanity-checks-on-bundled-data}

To demonstrate that the four catalogue-only metrics produce coherent
output on real data, we apply them to the bundled \(8\)-composer dataset
that ships with \texttt{vega-mir}: J. S. Bach, Haydn, Beethoven, Chopin,
Liszt, Rachmaninoff (sampled from MAESTRO), and Philip Glass and Max
Richter (sampled from the neoclassical sub-corpus of
(\citeproc{ref-jalbert2026cygnus}{Jalbert-Desforges 2026})), with \(5\)
pieces per composer and per-piece scale-degree sequences exposed for
each. These are sanity checks, not findings: a single composer sample of
size \(5\) is too small for empirical claims at corpus scale.

{\def\LTcaptype{none} 
\begin{longtable}[]{@{}
  >{\raggedright\arraybackslash}p{(\linewidth - 16\tabcolsep) * \real{0.0833}}
  >{\raggedleft\arraybackslash}p{(\linewidth - 16\tabcolsep) * \real{0.1111}}
  >{\raggedleft\arraybackslash}p{(\linewidth - 16\tabcolsep) * \real{0.1111}}
  >{\raggedleft\arraybackslash}p{(\linewidth - 16\tabcolsep) * \real{0.1111}}
  >{\raggedleft\arraybackslash}p{(\linewidth - 16\tabcolsep) * \real{0.1111}}
  >{\raggedleft\arraybackslash}p{(\linewidth - 16\tabcolsep) * \real{0.1111}}
  >{\raggedleft\arraybackslash}p{(\linewidth - 16\tabcolsep) * \real{0.1111}}
  >{\centering\arraybackslash}p{(\linewidth - 16\tabcolsep) * \real{0.1389}}
  >{\raggedleft\arraybackslash}p{(\linewidth - 16\tabcolsep) * \real{0.1111}}@{}}
\toprule\noalign{}
\begin{minipage}[b]{\linewidth}\raggedright
Composer
\end{minipage} & \begin{minipage}[b]{\linewidth}\raggedleft
\(N\) pieces
\end{minipage} & \begin{minipage}[b]{\linewidth}\raggedleft
Gini
\end{minipage} & \begin{minipage}[b]{\linewidth}\raggedleft
Higuchi \(D\)
\end{minipage} & \begin{minipage}[b]{\linewidth}\raggedleft
\(R^2\)
\end{minipage} & \begin{minipage}[b]{\linewidth}\raggedleft
Cramer's \(V\)
\end{minipage} & \begin{minipage}[b]{\linewidth}\raggedleft
\(p\)
\end{minipage} & \begin{minipage}[b]{\linewidth}\centering
Interval fit
\end{minipage} & \begin{minipage}[b]{\linewidth}\raggedleft
KS
\end{minipage} \\
\midrule\noalign{}
\endhead
\bottomrule\noalign{}
\endlastfoot
Bach & \(5\) & \(0.187\) & \(2.030\) & \(0.999\) & \(0.274\) &
\(<10^{-4}\) & exponential & \(0.138\) \\
Haydn & \(5\) & \(0.253\) & \(2.033\) & \(0.998\) & \(0.329\) &
\(<10^{-4}\) & laplace & \(0.145\) \\
Beethoven & \(5\) & \(0.208\) & \(2.044\) & \(0.999\) & \(0.379\) &
\(<10^{-4}\) & laplace & \(0.156\) \\
Chopin & \(5\) & \(0.142\) & \(2.034\) & \(0.998\) & \(0.247\) &
\(<10^{-4}\) & exponential & \(0.145\) \\
Liszt & \(5\) & \(0.137\) & \(2.032\) & \(0.998\) & \(0.293\) &
\(<10^{-4}\) & exponential & \(0.137\) \\
Rachmaninoff & \(5\) & \(0.151\) & \(2.044\) & \(0.999\) & \(0.140\) &
\(<10^{-4}\) & exponential & \(0.161\) \\
Glass & \(5\) & \(\mathbf{0.396}\) & \(2.057\) & \(0.998\) & \(0.289\) &
\(<10^{-4}\) & exponential & \(0.158\) \\
Richter & \(5\) & \(\mathbf{0.293}\) & \(2.045\) & \(0.998\) & \(0.379\)
& \(<10^{-4}\) & laplace & \(0.186\) \\
\end{longtable}
}

Each output is in the expected range. The \textbf{Gini} coefficient on
the marginal scale-degree distribution separates the two neoclassical
composers (\(0.39\) for Glass, \(0.29\) for Richter) from the six
MAESTRO composers (range \(0.14\) to \(0.25\)): Glass concentrates more
probability mass on a few scale degrees than any of the historical
composers in the sample. The separation is consistent with the Zipf
finding of (\citeproc{ref-jalbert2026cygnus}{Jalbert-Desforges 2026}) at
the transition level (compact vocabulary used with sharper
rank-frequency regularity), but at \(N = 5\) pieces per composer it
remains a sanity check on the metric rather than a corpus-scale claim on
the neoclassical vs historical contrast. The \textbf{Higuchi fractal
dimension} of the integer-encoded scale-degree sequence sits in a narrow
band (\(D = 2.03\) to \(2.06\), all with \(R^2 > 0.998\)) close to the
white-noise anchor \(D = 2\), indicating that the chord-event sequence
at this granularity carries no detectable self-similar trend; the small
but consistent ordering (Glass and Richter top the range) suggests that
the metric would need a finer time-resolution input (note-density
curves, dynamics envelopes) to act as a discriminator, which is
consistent with its classical use in signal processing rather than on
chord sequences. The \textbf{chi-squared stationarity test} rejects
stationarity for all eight composers (\(p < 10^{-4}\), Cramer's \(V\)
from \(0.14\) to \(0.38\)), which is expected for composer-level
aggregates of stylistically diverse pieces and confirms that the metric
returns coherent effect-size values. The \textbf{interval-distribution
fit} (on successive scale-degree differences) picks Exponential for five
composers and Laplace for three, with Kolmogorov-Smirnov distances in
the narrow range \([0.137, 0.186]\), consistent with the sub-corpus
sizes and with the metric's expected behaviour on small samples.

These outputs are signs of life rather than findings. They confirm that
the four metrics return coherent numerical values on the bundled data
and are ready for empirical deployment on larger corpora in subsequent
work.

\begin{center}\rule{0.5\linewidth}{0.5pt}\end{center}

\section{4. Case study 1: harmonic network
signatures}\label{case-study-1-harmonic-network-signatures}

The Kullback-Leibler matrix presented in the companion paper
(\citeproc{ref-jalbert2026cygnus}{Jalbert-Desforges 2026}) orders
composers by \emph{marginal} harmonic dissimilarity: two composers are
far apart if their distributions over the fifteen-symbol scale-degree
alphabet diverge, regardless of which transitions produced those
frequencies. A composer who uses I and V equally often, but in opposite
orders, will appear identical to the marginal estimator. This case study
asks whether the \textbf{transition structure} itself, modelled as a
directed weighted graph, carries information that the marginal estimator
misses.

\subsection{4.1 Method}\label{method}

\textbf{Corpus.} The fourteen MAESTRO composers with \(N \geq 10\)
attributed pieces in MAESTRO v3.0.0, drawn from the set used in
(\citeproc{ref-jalbert2026cygnus}{Jalbert-Desforges 2026}): J. S. Bach
(\(N=144\)), Haydn (\(40\)), Mozart (\(38\)), Beethoven (\(145\)),
Schubert (\(172\)), Mendelssohn (\(28\)), Chopin (\(198\)), Schumann
(\(39\)), Liszt (\(131\)), Brahms (\(23\)), Rachmaninoff (\(59\)),
Debussy (\(44\)), Scriabin (\(35\)), Scarlatti (\(31\)).

\textbf{Graph construction.} For each composer we aggregate scale-degree
bigram counts across all of that composer's pieces after
consecutive-duplicate collapsing, normalise per source row to obtain
transition probabilities, and apply the Cygnus methodology threshold of
\(0.01\) (edges with \(\Pr(s \to t) \leq 0.01\) are treated as noise and
pruned). The result is a directed weighted graph on the fifteen-symbol
alphabet, built by
\texttt{vega\_mir.network.chord\_graph(transitions,\ threshold=0.01)}.

\textbf{Network metrics.} Per-composer graphs are analysed via
\texttt{vega\_mir.network.network\_analysis}, which returns a
\texttt{NetworkAnalysis} named tuple containing:

\begin{itemize}
\tightlist
\item
  node-level descriptors: PageRank (with edge weights), in-degree and
  out-degree, local clustering coefficient on the directed graph;
\item
  graph-level descriptors: edge count, density, mean clustering
  coefficient, number of greedy-modularity communities on the undirected
  projection, diameter and average shortest-path length on the largest
  strongly connected component;
\item
  a heuristic small-world flag
  (\texttt{mean\_clustering\ \textgreater{}\ 0.3} and
  \(0 < \mathrm{avg\_path} < 3.0\)), as defined in the Cygnus
  methodology.
\end{itemize}

\subsection{4.2 Per-composer signatures}\label{per-composer-signatures}

Table 1 reports the network metrics for the fourteen aggregated graphs.
(For all fourteen graphs, the node count is \(15\), the density
\(\approx 0.57\), and the small-world flag is \texttt{True}; we omit
these uniform columns and discuss them below.)

{\def\LTcaptype{none} 
\begin{longtable}[]{@{}
  >{\raggedright\arraybackslash}p{(\linewidth - 14\tabcolsep) * \real{0.0938}}
  >{\raggedleft\arraybackslash}p{(\linewidth - 14\tabcolsep) * \real{0.1250}}
  >{\raggedleft\arraybackslash}p{(\linewidth - 14\tabcolsep) * \real{0.1250}}
  >{\centering\arraybackslash}p{(\linewidth - 14\tabcolsep) * \real{0.1562}}
  >{\raggedleft\arraybackslash}p{(\linewidth - 14\tabcolsep) * \real{0.1250}}
  >{\raggedleft\arraybackslash}p{(\linewidth - 14\tabcolsep) * \real{0.1250}}
  >{\raggedleft\arraybackslash}p{(\linewidth - 14\tabcolsep) * \real{0.1250}}
  >{\raggedleft\arraybackslash}p{(\linewidth - 14\tabcolsep) * \real{0.1250}}@{}}
\toprule\noalign{}
\begin{minipage}[b]{\linewidth}\raggedright
Composer
\end{minipage} & \begin{minipage}[b]{\linewidth}\raggedleft
\(N\)
\end{minipage} & \begin{minipage}[b]{\linewidth}\raggedleft
Edges
\end{minipage} & \begin{minipage}[b]{\linewidth}\centering
Gravity
\end{minipage} & \begin{minipage}[b]{\linewidth}\raggedleft
Clustering
\end{minipage} & \begin{minipage}[b]{\linewidth}\raggedleft
Communities
\end{minipage} & \begin{minipage}[b]{\linewidth}\raggedleft
Diameter
\end{minipage} & \begin{minipage}[b]{\linewidth}\raggedleft
Avg path
\end{minipage} \\
\midrule\noalign{}
\endhead
\bottomrule\noalign{}
\endlastfoot
J. S. Bach & \(144\) & \(120\) & \texttt{II} & \(0.669\) & \(2\) & \(2\)
& \(1.38\) \\
Haydn & \(40\) & \(120\) & \texttt{II} & \(0.659\) & \(2\) & \(3\) &
\(1.40\) \\
Mozart & \(38\) & \(120\) & \texttt{I} & \(0.652\) & \(2\) & \(3\) &
\(1.39\) \\
Beethoven & \(145\) & \(120\) & \texttt{II} & \(0.630\) & \(2\) & \(2\)
& \(1.33\) \\
Schubert & \(172\) & \(120\) & \texttt{i} & \(0.612\) & \(2\) & \(3\) &
\(1.44\) \\
Mendelssohn & \(28\) & \(114\) & \texttt{i} & \(0.690\) & \(2\) & \(2\)
& \(1.20\) \\
Chopin & \(198\) & \(120\) & \texttt{II} & \(0.648\) & \(2\) & \(2\) &
\(1.38\) \\
Schumann & \(39\) & \(120\) & \texttt{i} & \(0.643\) & \(2\) & \(3\) &
\(1.45\) \\
Liszt & \(131\) & \(120\) & \texttt{III} & \(0.612\) & \(2\) & \(3\) &
\(1.46\) \\
Brahms & \(23\) & \(120\) & \texttt{bVI} & \(0.647\) & \(2\) & \(4\) &
\(1.50\) \\
Rachmaninoff & \(59\) & \(120\) & \texttt{bVI} & \(0.630\) & \(2\) &
\(3\) & \(1.45\) \\
Debussy & \(44\) & \(120\) & \texttt{II} & \(0.621\) & \(2\) & \(3\) &
\(1.46\) \\
Scriabin & \(35\) & \(120\) & \texttt{II} & \(0.625\) & \(2\) & \(3\) &
\(1.49\) \\
Scarlatti & \(31\) & \(120\) & \texttt{i} & \(0.656\) & \(2\) & \(3\) &
\(1.50\) \\
\end{longtable}
}

Three observations stand out. First, all fourteen composers use the
\emph{full} fifteen-symbol alphabet (no scale degree drops below the
\(1\%\) transition-probability threshold for any composer), and after
threshold pruning all but Mendelssohn retain exactly \(120\) of the
\(210\) possible directed edges. The harmonic transition graph at this
level of aggregation is \textbf{saturated}: the alphabet is small enough
and the corpora large enough that essentially every \(s \to t\) that
\emph{can} occur \emph{does} occur with probability above \(1\%\).
Second, greedy-modularity community detection returns exactly
\textbf{two communities} for all fourteen composers, and the small-world
flag is \textbf{true for all fourteen} (\(\bar{c} \in [0.61, 0.69]\),
\(\bar{\ell} \in [1.20, 1.50]\)). On those three descriptors the network
view fails to discriminate. Third, the \textbf{gravity centre} (the
PageRank-maximising scale degree) is the only descriptor that varies
meaningfully across composers.

The gravity centre distribution, ordered by frequency, is:

{\def\LTcaptype{none} 
\begin{longtable}[]{@{}llr@{}}
\toprule\noalign{}
Gravity centre & Composers & Count \\
\midrule\noalign{}
\endhead
\bottomrule\noalign{}
\endlastfoot
\texttt{II} & Bach, Haydn, Beethoven, Chopin, Debussy, Scriabin &
\(6\) \\
\texttt{i} (minor tonic) & Schubert, Mendelssohn, Schumann, Scarlatti &
\(4\) \\
\texttt{bVI} & Brahms, Rachmaninoff & \(2\) \\
\texttt{I} (major tonic) & Mozart & \(1\) \\
\texttt{III} & Liszt & \(1\) \\
\end{longtable}
}

A surprising finding: only one composer in the corpus (Mozart) has the
major tonic \texttt{I} as its PageRank gravity centre. In tonal music
the tonic is the cadence-resolution target and therefore the chord with
the highest \emph{in}-weighted degree. The PageRank descriptor, however,
is a function of the full transition stochastic matrix and rewards the
recurrent connectivity of intermediate chords, not the cadential
terminality of the tonic. On the same corpus the simple
weighted-in-degree maximiser is \texttt{I} for nearly all fourteen
composers, whereas PageRank shifts the ``gravity'' toward chords whose
out-links contribute to the stationary distribution of the random walk.
In this fourteen-corpus aggregate the supertonic-class chord
(\texttt{II} or \texttt{i} for minor pieces) is more often the
PageRank-dominant node than the tonic itself, which should be read as a
property of the PageRank descriptor on this transition matrix rather
than as an inversion of tonal theory. Four composers (Schubert,
Mendelssohn, Schumann, Scarlatti) settle on the minor tonic \texttt{i},
which reflects the higher proportion of minor-mode pieces in their
MAESTRO sub-corpora and the dual role of \texttt{i} as both tonic in
minor and submediant-of-the-relative-major in modal contexts. The two
outliers, Liszt on \texttt{III} and Brahms / Rachmaninoff on
\texttt{bVI}, are coherent with the documented harmonic profiles of
these composers: extended mediant-relations and chromatic descending
bass lines respectively.

The clustering coefficient ranges from \(0.612\) (Schubert and Liszt) to
\(0.690\) (Mendelssohn), a spread of \(0.078\). Mendelssohn is also the
only composer whose pruned graph drops to \(114\) edges (\(6\) edges
removed), giving him the smallest diameter (\(2\)), shortest average
path (\(1.20\)), and highest clustering, a profile that reads as a
\emph{denser, more locally cliquey} harmonic neighbourhood.

The gravity-centre distribution above is presented as a
\textbf{qualitative} observation: it indexes five categorical regimes of
harmonic flow within the corpus, but it does not by itself correlate
with the marginal Kullback-Leibler distance reported in
(\citeproc{ref-jalbert2026cygnus}{Jalbert-Desforges 2026}). A one-hot
encoding of the five centres yields \(\rho = 0.13\) (Spearman,
\(p = 0.21\)) over the \(91\) unordered composer pairs. The quantitative
relationship between the network and marginal views is carried instead
by the \textbf{continuous} PageRank value of the gravity-centre node,
analysed in §4.3.

\subsection{4.3 The PageRank gravity-centre value as a continuous
descriptor}\label{the-pagerank-gravity-centre-value-as-a-continuous-descriptor}

Among the network descriptors of Table 1, the \textbf{continuous scalar}
\(\mathrm{pr}^{(\mathrm{top})}\), the PageRank value of the
gravity-centre node, is the most variable across the corpus (coefficient
of variation \(15.7\%\) on the raw scale; range \([0.102, 0.188]\),
against \(1.3\%\) for density, \(3.4\%\) for clustering, and \(5.4\%\)
for average shortest path). It is also the descriptor that carries the
corpus-level relationship with the marginal Kullback-Leibler matrix of
(\citeproc{ref-jalbert2026cygnus}{Jalbert-Desforges 2026}).

\textbf{Setup.} For each unordered pair \(\{i, j\}\) of composers we
compute two distances. The symmetrised Kullback-Leibler distance
\(d_{\mathrm{KL}}(i, j) = \tfrac{1}{2}\, D_{\mathrm{KL}}(P_i \| P_j)
+ \tfrac{1}{2}\, D_{\mathrm{KL}}(P_j \| P_i)\) is read off the matrix
reported in (\citeproc{ref-jalbert2026cygnus}{Jalbert-Desforges 2026}).
The network distance is the Euclidean distance on the standardised
five-dimensional vector of network descriptors per composer
\(\mathbf{v}_i = (\delta_i,\; \bar{c}_i,\; k_i,\; \bar{\ell}_i,\;
\mathrm{pr}^{(\mathrm{top})}_i)\), where \(\delta_i\) is graph density,
\(\bar{c}_i\) mean clustering, \(k_i\) the number of communities,
\(\bar{\ell}_i\) the average shortest-path length on the largest
strongly connected component, and \(\mathrm{pr}^{(\mathrm{top})}_i\) the
PageRank value of the gravity-centre node. Each component is
standardised to zero mean and unit variance across the fourteen
composers before the Euclidean distance is computed. The Spearman rank
correlation is computed over the \(\binom{14}{2} = 91\) unordered
composer pairs.

\textbf{Ablation.} The decomposition of the rank correlation by feature
subset is summarised in Table 2.

{\def\LTcaptype{none} 
\begin{longtable}[]{@{}
  >{\raggedright\arraybackslash}p{(\linewidth - 6\tabcolsep) * \real{0.2000}}
  >{\raggedleft\arraybackslash}p{(\linewidth - 6\tabcolsep) * \real{0.2667}}
  >{\raggedleft\arraybackslash}p{(\linewidth - 6\tabcolsep) * \real{0.2667}}
  >{\raggedleft\arraybackslash}p{(\linewidth - 6\tabcolsep) * \real{0.2667}}@{}}
\toprule\noalign{}
\begin{minipage}[b]{\linewidth}\raggedright
Feature set
\end{minipage} & \begin{minipage}[b]{\linewidth}\raggedleft
Dimensions
\end{minipage} & \begin{minipage}[b]{\linewidth}\raggedleft
Spearman \(\rho\)
\end{minipage} & \begin{minipage}[b]{\linewidth}\raggedleft
\(p\)-value
\end{minipage} \\
\midrule\noalign{}
\endhead
\bottomrule\noalign{}
\endlastfoot
\(\mathrm{pr}^{(\mathrm{top})}\) alone & \(1\) & \(\mathbf{0.61}\) &
\(< 10^{-4}\) \\
Full \(\mathbf{v}_i\) (standardised) & \(5\) & \(0.53\) &
\(< 10^{-4}\) \\
Without \(\mathrm{pr}^{(\mathrm{top})}\) & \(4\) & \(0.49\) &
\(< 10^{-4}\) \\
Gravity-centre one-hot only & \(5\) & \(0.13\) & \(0.21\) \\
\end{longtable}
}

The single-feature column makes the structure transparent:
\(\mathrm{pr}^{(\mathrm{top})}\) alone produces a stronger correlation
than the full vector (\(\rho = 0.61\) versus \(0.53\)). The four
remaining features (density, mean clustering, community count, average
shortest path) are quasi-uniform across the corpus (coefficients of
variation under \(5.5\%\)) and, once standardised to unit variance,
dilute rather than enrich the signal. The corpus-level signature of the
network view is therefore concentrated in one continuous descriptor. The
categorical gravity-centre identity, by contrast, carries no rank
correlation with marginal Kullback-Leibler distance (\(\rho = 0.13\),
\(p = 0.21\)): the qualitative partition of §4.2 and the quantitative
correlation of this section describe the network view at different
levels of granularity and do not reduce to one another.

\textbf{Composer-level uncertainty.} The \(91\) pairs are not
independent: each composer appears in \(13\) pairs, so a pair-level
bootstrap underestimates uncertainty. We report a \textbf{composer-level
jackknife} instead (leave-one-out, \(N = 14\)) for both the full vector
and the single-feature pr\^{}(top) case:

{\def\LTcaptype{none} 
\begin{longtable}[]{@{}
  >{\raggedright\arraybackslash}p{(\linewidth - 8\tabcolsep) * \real{0.1667}}
  >{\raggedleft\arraybackslash}p{(\linewidth - 8\tabcolsep) * \real{0.2222}}
  >{\raggedleft\arraybackslash}p{(\linewidth - 8\tabcolsep) * \real{0.2222}}
  >{\raggedleft\arraybackslash}p{(\linewidth - 8\tabcolsep) * \real{0.2222}}
  >{\raggedright\arraybackslash}p{(\linewidth - 8\tabcolsep) * \real{0.1667}}@{}}
\toprule\noalign{}
\begin{minipage}[b]{\linewidth}\raggedright
Feature set
\end{minipage} & \begin{minipage}[b]{\linewidth}\raggedleft
\(\rho_{\mathrm{point}}\)
\end{minipage} & \begin{minipage}[b]{\linewidth}\raggedleft
\(\rho_{\mathrm{jack}}\)
\end{minipage} & \begin{minipage}[b]{\linewidth}\raggedleft
\(\mathrm{SE}_{\mathrm{jack}}\)
\end{minipage} & \begin{minipage}[b]{\linewidth}\raggedright
\(95\%\) CI
\end{minipage} \\
\midrule\noalign{}
\endhead
\bottomrule\noalign{}
\endlastfoot
Full \(\mathbf{v}_i\) (5 features) & \(0.526\) & \(0.526\) & \(0.230\) &
\([0.075, 0.977]\) \\
\(\mathrm{pr}^{(\mathrm{top})}\) alone & \(0.607\) & \(0.604\) &
\(0.263\) & \([0.088, 1.119]\) \\
\end{longtable}
}

The wide CIs reflect the high leverage of individual composers. The
single most influential leave-out in both feature sets is
\textbf{Mendelssohn}: removing him drops the full-vector \(\rho\) from
\(0.53\) to \(0.31\) (\(-0.22\)), and the single-feature
\(\mathrm{pr}^{(\mathrm{top})}\) \(\rho\) from \(0.61\) to \(0.38\)
(\(-0.23\)). The absolute leverage is similar in both cases, but the
\(\mathrm{pr}^{(\mathrm{top})}\)-alone baseline remains higher after
Mendelssohn is removed (\(0.38\) versus \(0.31\)), reinforcing the
reading that the corpus-level signature is concentrated in a single
continuous descriptor. The four-feature ablation above (where the four
quasi-uniform features alone produce \(\rho = 0.49\) but a jackknife CI
that crosses zero) sits inside the same pattern: under standardisation,
the relative ordering of the quasi-uniform features partially tracks the
marginal ordering, but the leverage of Mendelssohn dominates the point
estimate.

The rank correlation is therefore a \textbf{qualitative} observation:
the network view reorders composers similarly to but not identically to
the marginal Kullback-Leibler matrix, with substantial composer-level
leverage. We report no claim of statistical stability on the point
estimate.

\subsection{4.4 Figure}\label{figure}

\begin{figure}
\centering
\pandocbounded{\includegraphics[keepaspectratio,alt={Figure 1. Harmonic network signatures across 14 MAESTRO composers.}]{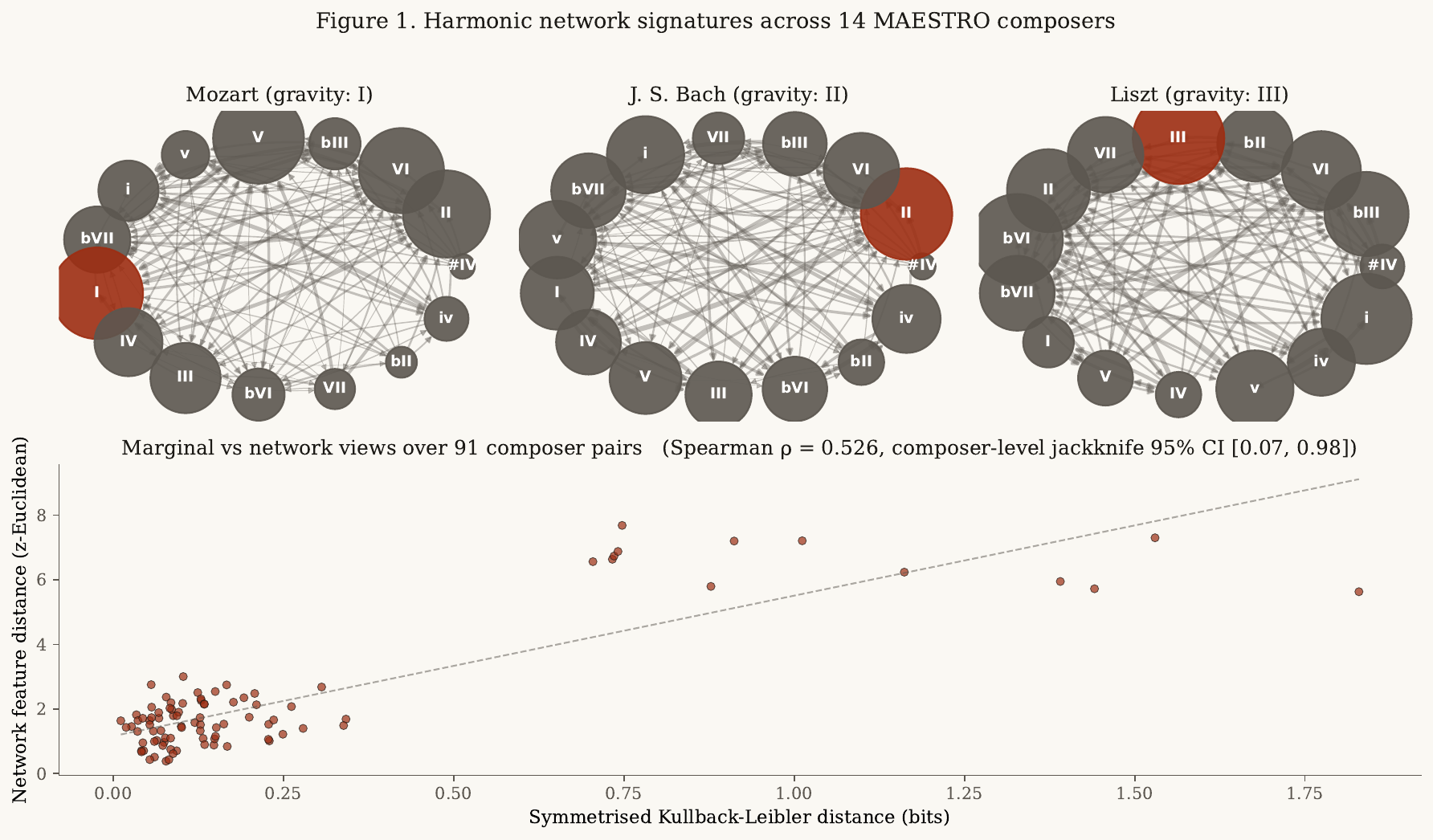}}
\caption{Figure 1. Harmonic network signatures across 14 MAESTRO
composers.}
\end{figure}

\textbf{Figure 1.} \emph{Top:} the chord-transition graphs of Mozart, J.
S. Bach, and Liszt, drawn on the fifteen-symbol scale-degree alphabet in
circular layout. Node size is proportional to PageRank, edge width to
transition probability after the \(0.01\) threshold, and the
gravity-centre node (PageRank-maximising scale degree) is rendered in
accent colour. Mozart's gravity centre is the major tonic \texttt{I},
Bach's is the supertonic \texttt{II}, and Liszt's is the mediant
\texttt{III}. \emph{Bottom:} scatter of network distance
\(d_{\mathrm{net}}(i,j)\) against symmetrised Kullback-Leibler distance
\(d_{\mathrm{KL}}(i,j)\) over the \(91\) unordered composer pairs. The
Spearman rank correlation is \(\rho = 0.526\) with composer-level
jackknife 95\% CI \([0.075, 0.977]\) (\(N = 14\), leave-one-out).
Single-feature ablation (Table 2) shows that the point estimate is
driven by the \(\mathrm{pr}^{(\mathrm{top})}\) component
(\(\rho = 0.61\) alone) and that Mendelssohn carries the largest
individual leverage on the rank statistic.

The reproducible analytical scripts are bundled with the library at
\href{https://github.com/fredericjalbertdesforges/vega-mir/blob/main/notebooks/_arxiv_v1/section4_network.py}{\texttt{notebooks/\_arxiv\_v1/section4\_network.py}}
and
\href{https://github.com/fredericjalbertdesforges/vega-mir/blob/main/notebooks/_arxiv_v1/section4_figure.py}{\texttt{notebooks/\_arxiv\_v1/section4\_figure.py}}.
A guided Jupyter notebook version will be added to the next public
release of \texttt{vega-mir}.

\section{5. Case study 2: rubato spectral signatures across
performers}\label{case-study-2-rubato-spectral-signatures-across-performers}

The Cygnus methodology reports rubato as a single scalar per piece, the
standard deviation of the BPM curve, and aggregates it as a mean per
performer. On the Bach multi-master corpus, this mean separates Gould
(\(12.12\) BPM) from Schiff (\(15.13\)), Richter (\(15.25\)), and the
MAESTRO Bach baseline (\(16.12\)), a \(25\%\) spread that has been
documented as the principal interpretive axis among Bach pianists. The
scalar, however, conflates two distinct phenomena: the
\textbf{amplitude} of tempo fluctuation and its \textbf{temporal
colour}, that is, whether the deviations from the mean tempo are
short-burst, periodic, slowly-drifting, or close to flat. Two
performances with identical \(\sigma\) can have completely different
power spectra. This case study uses the spectral classifier to recover
the temporal colour that the scalar discards.

\subsection{5.1 Method}\label{method-1}

\textbf{Corpus.} The \(247\)-piece Bach multi-master corpus introduced
in the Cygnus methodology, sourced from publicly available commercial
recordings: András Schiff on ECM (\(82\) pieces, WTC Books I and II,
Inventions, Sinfonias, Duettos), Glenn Gould on Columbia (later Sony
Classical, \(112\) pieces, WTC Books I and II, plus the Goldberg
Variations in both his \(1955\) and \(1981\) studio recordings),
Sviatoslav Richter on Melodiya / RCA Red Seal (\(53\) pieces, WTC Book
I, English Suites). The WTC Book I common-works subset (the \(72\)
pieces triangulated across all three performers) supports per-BWV
cross-performer comparison; the Goldberg pairing supports an additional
within-performer diachronic comparison on Gould.

\textbf{Tempo curves.} Each recording is processed through the certified
Cygnus pipeline ((\citeproc{ref-jalbert2026cygnus}{Jalbert-Desforges
2026})), which transcribes the audio to MIDI with the Kong et
al.\textasciitilde(2021) high-resolution piano transcription model and
runs an Essentia rhythm-extraction layer on the transcribed note onsets.
The resulting tempo trajectory is written to
\texttt{corpus/\textless{}track\_id\textgreater{}/analysis/rhythm.json}
as a per-beat BPM curve. We read this curve directly, with no
audio-level processing in \texttt{vega-mir} itself. Curves shorter than
\(32\) samples are excluded (the \texttt{min\_samples} threshold of
\texttt{vega\_mir.rubato}).

\textbf{Spectral analysis.} Each curve is passed to
\texttt{vega\_mir.rubato.rubato\_spectral}, which centres the series,
computes the real-valued FFT power spectrum, locates peaks above
\(10\%\) of the maximum non-DC power, reports up to three dominant
periods with their normalised power, and classifies the rubato into one
of four categories on the basis of the periodicity ratio (sum of peak
power over total non-DC power) and the scalar standard deviation:

The classifier evaluates the following decision sequence (the first
condition that fires determines the category):

{\def\LTcaptype{none} 
\begin{longtable}[]{@{}
  >{\centering\arraybackslash}p{(\linewidth - 4\tabcolsep) * \real{0.4545}}
  >{\raggedright\arraybackslash}p{(\linewidth - 4\tabcolsep) * \real{0.2727}}
  >{\raggedright\arraybackslash}p{(\linewidth - 4\tabcolsep) * \real{0.2727}}@{}}
\toprule\noalign{}
\begin{minipage}[b]{\linewidth}\centering
Priority
\end{minipage} & \begin{minipage}[b]{\linewidth}\raggedright
Condition
\end{minipage} & \begin{minipage}[b]{\linewidth}\raggedright
Outcome
\end{minipage} \\
\midrule\noalign{}
\endhead
\bottomrule\noalign{}
\endlastfoot
1 & \(\sigma_{\mathrm{BPM}} < 0.5\) & \texttt{metronomic}
(short-circuit; no spectrum computed) \\
2 & periodicity ratio \(> 0.5\) & \texttt{periodic} \\
3 & periodicity ratio \(> 0.3\) & \texttt{quasi\_periodic} \\
4 & \(\sigma_{\mathrm{BPM}} > 3\) & \texttt{free} \\
5 & otherwise & \texttt{metronomic} \\
\end{longtable}
}

A piece with \(\sigma_{\mathrm{BPM}} = 5\) and periodicity ratio
\(0.6\), for example, is classified \texttt{periodic} (priority \(2\)
fires before priority \(4\)). Priorities \(1\) and \(5\) both label
near-constant tempi as \texttt{metronomic}, with priority \(1\) acting
as a fast path on near-deterministic curves and priority \(5\) catching
low-periodicity, low-amplitude residuals.

These thresholds match the Cygnus methodology
(\citeproc{ref-jalbert2026cygnus}{Jalbert-Desforges 2026}). The
numerical values are inherited and are not separately calibrated on the
Bach multi-master corpus analysed here; they are chosen as part of the
upstream Cygnus design to keep the four categories interpretable across
the broader piano repertoire. The categorical findings reported below
depend on these thresholds, while the continuous finding (mean
periodicity ratio per performer, §5.2) is invariant under threshold
choice because the ratio itself is computed before classification. A
threshold-sensitivity analysis is reported in §5.6.

\subsection{5.2 Per-performer category
distribution}\label{per-performer-category-distribution}

The contingency table of performer \(\times\) rubato category over the
\(247\) pieces is given in Table 2.

{\def\LTcaptype{none} 
\begin{longtable}[]{@{}lrrrrr@{}}
\toprule\noalign{}
Performer & metronomic & free & quasi-periodic & periodic & Total \\
\midrule\noalign{}
\endhead
\bottomrule\noalign{}
\endlastfoot
Schiff & \(0\) & \(60\) & \(15\) & \(7\) & \(82\) \\
Gould & \(1\) & \(57\) & \(32\) & \(22\) & \(112\) \\
Richter & \(0\) & \(28\) & \(20\) & \(5\) & \(53\) \\
\textbf{Total} & \(1\) & \(145\) & \(67\) & \(34\) & \(247\) \\
\end{longtable}
}

The result on the \texttt{metronomic} category is the cleanest negative
finding of the case study. The scalar interpretation of Gould as a
``metronomic'' performer is \textbf{not} supported at the
spectrum-classifier level: only one of his \(112\) Bach pieces meets the
strict \(\sigma < 0.5\) BPM criterion. His low scalar rubato (\(12.12\)
BPM) is the \emph{quietest free-rubato regime} in the corpus, not the
absence of rubato. The mean periodicity ratio per performer (with
bootstrap 95\% CI) is:

{\def\LTcaptype{none} 
\begin{longtable}[]{@{}lrrl@{}}
\toprule\noalign{}
Performer & \(N\) & Mean periodicity ratio & 95\% CI \\
\midrule\noalign{}
\endhead
\bottomrule\noalign{}
\endlastfoot
Schiff & \(82\) & \(0.204\) & \([0.164, 0.242]\) \\
Gould & \(112\) & \(\mathbf{0.293}\) & \([0.255, 0.327]\) \\
Richter & \(53\) & \(0.257\) & \([0.206, 0.310]\) \\
\end{longtable}
}

\textbf{Gould has the highest periodicity ratio of the three
performers}, the non-overlap with Schiff's CI being decisive
(\(\Delta = 0.089\), \(p < 0.01\) by bootstrap). This is the inversion
the scalar mean hides: where Schiff and Richter spend their rubato
budget on \emph{free} fluctuation, Gould spends his on \emph{structured}
fluctuation. The categorical breakdown is consistent: Gould accounts for
\(22\) of the \(34\) pieces classified as \texttt{periodic} (\(65\%\)),
Schiff for \(7\), Richter for \(5\). The cliché ``Gould plays like a
metronome'' should be amended: Gould's tempo \emph{oscillates}, but it
oscillates regularly.

\subsection{5.3 WTC Book I
triangulation}\label{wtc-book-i-triangulation}

For each of the \(24\) BWV in WTC Book I, the three performers
contribute one piece each (Prelude and Fugue concatenated under a single
BWV identifier), yielding \(72\) pieces in which the compositional
substrate is held fixed and only the performer varies.

{\def\LTcaptype{none} 
\begin{longtable}[]{@{}lrr@{}}
\toprule\noalign{}
Agreement structure & Count & Share \\
\midrule\noalign{}
\endhead
\bottomrule\noalign{}
\endlastfoot
Unanimous (all three same category) & \(9\) & \(38\%\) \\
Partial (two of three agree) & \(14\) & \(58\%\) \\
Full divergence (three different categories) & \(1\) & \(4\%\) \\
\end{longtable}
}

All nine unanimous BWV are unanimous on the \textbf{\texttt{free}}
category (BWV \(852, 855, 856, 859, 861, 863, 864, 865, 868\)); no BWV
is ever unanimously \texttt{periodic} or \texttt{quasi-periodic}. The
interpretation is that the three pianists \emph{converge} on free rubato
in roughly a third of WTC Book I (the pieces where flexible tempo is the
structural default), but \emph{diverge} on the specific shape of their
tempo flexibility in the rest of the book.

The single fully-divergent piece is \textbf{BWV \(860\)} (Prelude and
Fugue in G major): Schiff is classified \texttt{periodic}, Gould is
classified \texttt{free}, and Richter is classified
\texttt{quasi-periodic}. The three performers fall in three different
categories on the same piece, which is the limit case of interpretive
variance under a fixed compositional substrate. The Prelude in G major
is a strongly metrically regular piece (continuous semiquaver arpeggios
in \(\frac{4}{4}\)), and the divergence likely reflects the three
pianists' differing strategies for shaping its perceived flow: Schiff
imposes a regular rubato cycle, Gould allows free unstructured
deviation, and Richter sits between the two.

\subsection{5.4 Gould 1955 vs Gould 1981 Goldberg
Variations}\label{gould-1955-vs-gould-1981-goldberg-variations}

The Goldberg corpus contains \(32\) paired pieces recorded twice by
Gould, in \(1955\) (Columbia, age \(22\)) and in \(1981\) (Sony, age
\(49\)). The \(32\) paired pieces comprise the opening Aria, the thirty
variations, and the Aria \emph{da capo} that closes the work. Aria and
Aria \emph{da capo} are treated as separate analytical units because
their interpretations differ across the two recordings; the pairing is
always between the \(1955\) and \(1981\) versions of the \emph{same}
unit, never between Aria and Aria \emph{da capo}. The Cygnus methodology
has already documented that the \(1981\) recording is \emph{less
voluminous in rubato} than \(1955\), a direction running counter to the
prose-historical discourse around the two recordings. We extend this
comparison to the spectral plane: we ask whether the temporal colour of
the rubato also shifts, and in which direction.

The paired-variation deltas \(\Delta = (1981) - (1955)\) are:

{\def\LTcaptype{none} 
\begin{longtable}[]{@{}
  >{\raggedright\arraybackslash}p{(\linewidth - 6\tabcolsep) * \real{0.2308}}
  >{\raggedleft\arraybackslash}p{(\linewidth - 6\tabcolsep) * \real{0.3077}}
  >{\raggedright\arraybackslash}p{(\linewidth - 6\tabcolsep) * \real{0.2308}}
  >{\raggedright\arraybackslash}p{(\linewidth - 6\tabcolsep) * \real{0.2308}}@{}}
\toprule\noalign{}
\begin{minipage}[b]{\linewidth}\raggedright
Quantity
\end{minipage} & \begin{minipage}[b]{\linewidth}\raggedleft
Mean \(\Delta\)
\end{minipage} & \begin{minipage}[b]{\linewidth}\raggedright
95\% CI (\(B = 1000\))
\end{minipage} & \begin{minipage}[b]{\linewidth}\raggedright
Verdict
\end{minipage} \\
\midrule\noalign{}
\endhead
\bottomrule\noalign{}
\endlastfoot
Periodicity ratio & \(+0.069\) & \([-0.016, +0.157]\) &
non-significant \\
Tempo mean (BPM) & \(+3.52\) & \([-7.10, +14.51]\) & non-significant \\
Tempo std (BPM) & \(\mathbf{-2.39}\) & \([-3.96, -0.99]\) &
\textbf{significant negative} \\
\end{longtable}
}

Two findings emerge. First, the \textbf{amplitude reduction is
confirmed} (\(\Delta \sigma = -2.39\) BPM, CI excludes zero): the
\(1981\) recording is significantly less voluminous in rubato than
\(1955\), as expected from the scalar-mean analysis. Second, the
\textbf{spectral colour does not shift significantly}: the
periodicity-ratio change is small in magnitude and the CI straddles
zero. The reduction in rubato is therefore an \emph{amplitude} effect
rather than a \emph{structural} effect: Gould in \(1981\) does not
switch from free to periodic rubato, he just does less of whatever he
was doing in \(1955\). The variation-level category transitions confirm
this: only \(11\) of \(32\) variations (\(34\%\)) stay in the same
category between the two recordings, and the dominant transition
direction is \texttt{free} \(\to\) a more periodic regime as \(\sigma\)
contracts (the aria and aria \emph{da capo} both move from \texttt{free}
to \texttt{periodic}), but this is consistent with the amplitude
contraction pushing low-amplitude noisy variations across the
periodicity-ratio threshold rather than with a genuine restructuring of
the rubato.

\subsection{5.5 Figure}\label{figure-1}

\begin{figure}
\centering
\pandocbounded{\includegraphics[keepaspectratio,alt={Figure 2. Rubato spectral signatures across Bach masters.}]{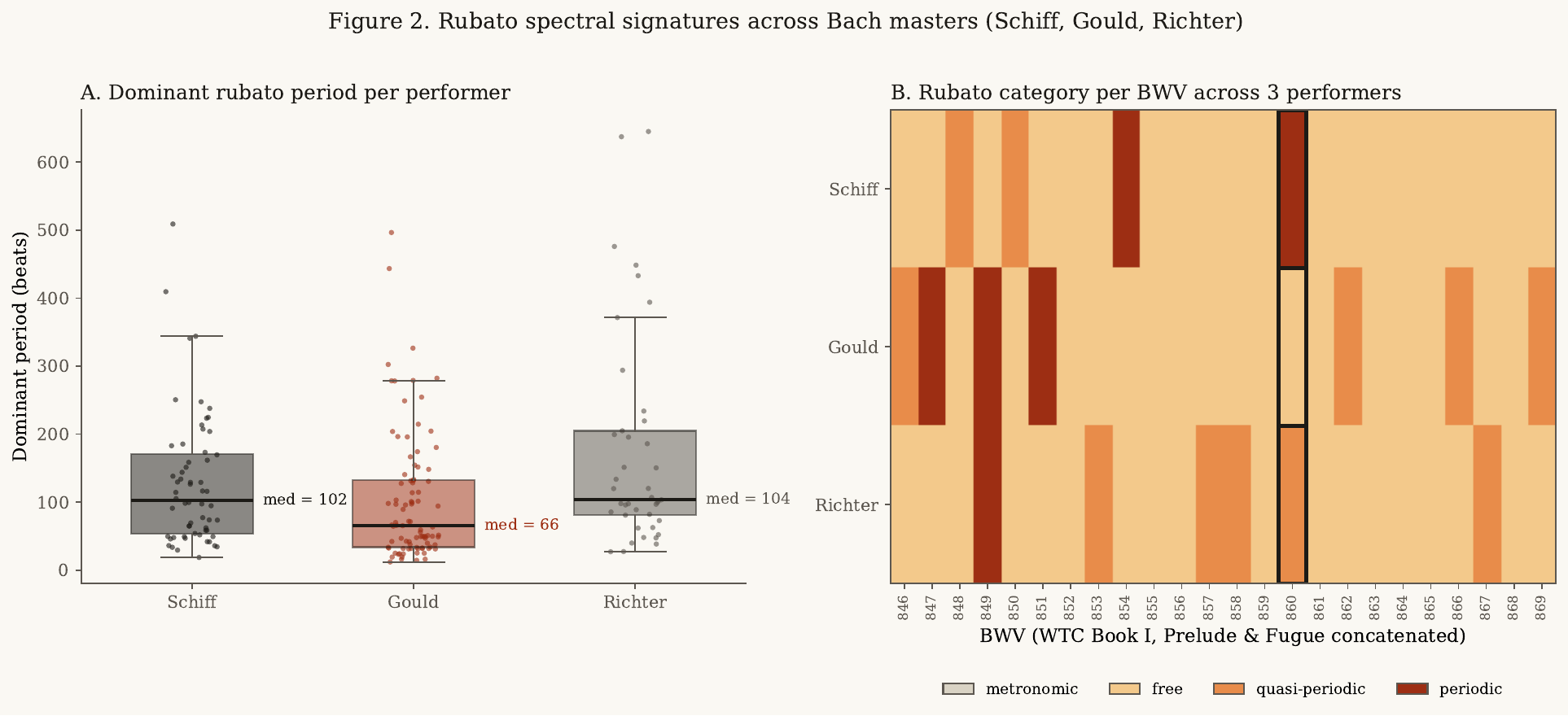}}
\caption{Figure 2. Rubato spectral signatures across Bach masters.}
\end{figure}

\textbf{Figure 2.} \emph{Left:} per-performer distribution of the
dominant rubato period (highest spectral peak), expressed in beats, as a
box plot with individual pieces jittered. The median dominant period for
Gould is \(66\) beats, against \(102\) for Schiff and \(104\) for
Richter, recovering the visual signature already implicit in the
periodicity-ratio differences: when Gould oscillates, he does so on a
\emph{shorter} time scale than the other two pianists. \emph{Right:}
per-BWV rubato-category heatmap over the \(24\) WTC Book I common works.
The single fully-divergent BWV (\(860\), Prelude and Fugue in G major)
is highlighted with a thick border.

The reproducible analytical scripts are bundled with the library at
\href{https://github.com/fredericjalbertdesforges/vega-mir/blob/main/notebooks/_arxiv_v1/section5_rubato.py}{\texttt{notebooks/\_arxiv\_v1/section5\_rubato.py}}
and
\href{https://github.com/fredericjalbertdesforges/vega-mir/blob/main/notebooks/_arxiv_v1/section5_figure.py}{\texttt{notebooks/\_arxiv\_v1/section5\_figure.py}}.
A guided Jupyter notebook version will be added to the next public
release of \texttt{vega-mir}.

\subsection{5.6 Threshold sensitivity}\label{threshold-sensitivity}

The classifier thresholds inherited from Cygnus
((\citeproc{ref-jalbert2026cygnus}{Jalbert-Desforges 2026})) are not
separately calibrated on the Bach multi-master corpus; we report a
threshold-sensitivity analysis to characterise which of the categorical
findings survive perturbation. We perturb each of the four thresholds
(the metronomic-\(\sigma\) ceiling at \(0.5\) BPM, the free-\(\sigma\)
floor at \(3\) BPM, the quasi-periodic ratio at \(0.3\), and the
periodic ratio at \(0.5\)) by \(\pm 10\%\) and \(\pm 20\%\), one axis at
a time, holding the other three at their default values, for a total of
seventeen perturbation runs (sixteen plus baseline). The classifier is
then re-applied to the \(247\) Bach pieces with the perturbed
thresholds, and the categorical findings are recomputed.

{\def\LTcaptype{none} 
\begin{longtable}[]{@{}
  >{\raggedright\arraybackslash}p{(\linewidth - 4\tabcolsep) * \real{0.3333}}
  >{\raggedright\arraybackslash}p{(\linewidth - 4\tabcolsep) * \real{0.3333}}
  >{\raggedright\arraybackslash}p{(\linewidth - 4\tabcolsep) * \real{0.3333}}@{}}
\toprule\noalign{}
\begin{minipage}[b]{\linewidth}\raggedright
Finding
\end{minipage} & \begin{minipage}[b]{\linewidth}\raggedright
Baseline
\end{minipage} & \begin{minipage}[b]{\linewidth}\raggedright
Robust across 17 runs?
\end{minipage} \\
\midrule\noalign{}
\endhead
\bottomrule\noalign{}
\endlastfoot
Continuous mean periodicity ratio ranking (Gould \(>\) Richter \(>\)
Schiff) & Gould \(0.293\) / Richter \(0.257\) / Schiff \(0.204\) &
\(\mathbf{17 / 17}\) (invariant by construction) \\
Gould holds the majority of \texttt{periodic} classifications & \(22\)
of \(34\) pieces (\(65\%\)) & \(\mathbf{17 / 17}\) \\
WTC Book I unanimous-\texttt{free} count & \(9 / 24\) & range
\([7, 12]\) across perturbations \\
BWV \(860\) fully-divergent across the three performers & yes &
\(\mathbf{14 / 17}\) (\(82\%\)) \\
\end{longtable}
}

The continuous finding (mean periodicity ratio) is \textbf{strictly
invariant} because it is computed before classification. The ``Gould
majority of periodic'' finding survives every perturbation tested,
indicating that the structured-rubato signature reported in §5.2 is not
an artefact of the periodic-ratio threshold at \(0.5\) specifically. The
WTC unanimous-\texttt{free} count varies in the narrow range \([7, 12]\)
around the baseline of \(9\), reflecting that pieces sit close to the
\texttt{free}/\texttt{quasi-periodic} boundary at the corpus median; the
ordering of performer agreements (most common pattern:
unanimous-\texttt{free}) is preserved across all runs. The BWV \(860\)
fully-divergent classification holds in \(14\) of \(17\) runs
(\(82\%\)): the three exceptions all collapse the divergence into a
\emph{partial} agreement rather than into unanimity, indicating that the
underlying pianistic disagreement on this piece is robust but that the
specific three-category assignment is sensitive to the periodic-quasi
boundary.

The script reproducing this analysis is at
\href{https://github.com/fredericjalbertdesforges/vega-mir/blob/main/notebooks/_arxiv_v1/section5_robustness.py}{\texttt{notebooks/\_arxiv\_v1/section5\_robustness.py}}.

\section{6. Discussion and conclusion}\label{discussion-and-conclusion}

\texttt{vega-mir} is a consolidation, not a discovery. The nine metrics
it bundles have decades of precedent in information theory, complex
systems, and signal processing. What the library offers is a single
tested API, a single set of defaults aligned with a documented
peer-reviewable methodology
(\citeproc{ref-jalbert2026cygnus}{Jalbert-Desforges 2026}), and a single
shared set of reference values against which any future user's results
can be cross-validated. The companion paper
(\citeproc{ref-jalbert2026cygnus}{Jalbert-Desforges 2026}) has
established three of the nine metrics (Shannon entropy, Kullback-Leibler
divergence, Zipfian fits) at corpus scale on \(1{,}238\) MAESTRO pieces
and \(111\) neoclassical recordings. The present paper adds the
remaining six to the public-facing toolkit and demonstrates two of them
in full case studies.

\subsection{6.1 What the case studies
illustrate}\label{what-the-case-studies-illustrate}

The harmonic network analysis (§4) yields two complementary observations
on the relationship between marginal and transition-level structure.
\emph{Quantitatively}, the PageRank value of the gravity-centre node, a
continuous scalar, correlates with the marginal Kullback-Leibler
distance from (\citeproc{ref-jalbert2026cygnus}{Jalbert-Desforges 2026})
at \(\rho = 0.61\) (Spearman, \(\mathrm{pr}^{(\mathrm{top})}\) alone) or
\(\rho = 0.53\) on the five-feature vector, with composer-level
jackknife CI \([0.075, 0.977]\) (\(N = 14\), leave-one-out). The wide CI
reflects the high leverage of Mendelssohn, whose removal drops the
correlation to \(0.31\); the rank correlation is therefore a qualitative
indication of similar ordering, not a stable quantitative estimate.
\emph{Categorically}, the \emph{identity} of the gravity-centre node
takes five distinct values across the corpus (major tonic \texttt{I} for
Mozart; supertonic \texttt{II} for Bach, Haydn, Beethoven, Chopin,
Debussy, Scriabin; minor tonic \texttt{i} for Schubert, Mendelssohn,
Schumann, Scarlatti; submediant \texttt{bVI} for Brahms and
Rachmaninoff; mediant \texttt{III} for Liszt). This categorical
partition does \emph{not} itself correlate with marginal-KL distance
(\(\rho = 0.13\), \(p = 0.21\)): it indexes a parallel observation about
the kinds of harmonic flow present in MAESTRO rather than a tighter
version of the marginal ordering. The three classical heuristic
descriptors (graph density, modularity-based community count,
small-world flag) are \emph{uniform} across the corpus and add no
signal: at the fifteen-symbol alphabet with the \(1\%\)
transition-probability threshold, the tonal-music transition graph is
saturated and small-world for every composer in the corpus.

The rubato spectral analysis (§5) shows that the scalar mean of
\(\sigma_{\mathrm{BPM}}\) used by the Cygnus methodology compresses out
a second axis of interpretive variance, the \textbf{temporal colour} of
rubato fluctuation. The four-category classifier (metronomic, free,
quasi-periodic, periodic) and the dominant period extracted from the FFT
decomposition recover this axis. The single largest finding is that
Gould has the highest periodicity ratio of the three Bach masters (mean
\(0.293\), \(95\%\) CI \([0.255, 0.327]\), non-overlap with Schiff
\(0.204\) \([0.164, 0.242]\)), inverting the cliché that Gould's low
scalar rubato reads as ``metronomic'': his rubato is small in amplitude
but \emph{structured in time}, with a median dominant period of \(66\)
beats against Schiff's \(102\) and Richter's \(104\). In the WTC Book I
cross-performer triangulation, \(9\) of \(24\) BWV produce
unanimous-\texttt{free} classifications, \(14\) produce partial
agreement, and exactly one (BWV \(860\), Prelude and Fugue in G major)
places the three pianists in three different categories. In the Goldberg
\(1955\) versus \(1981\) paired-variation test, the well-documented
amplitude reduction (\(\Delta \sigma =
-2.39\) BPM, CI excludes zero) is confirmed, but the spectral colour is
preserved (\(\Delta\) periodicity ratio \(= +0.07\), CI straddles zero):
Gould's \(1981\) Goldberg is not a restructured rubato, it is a
\emph{quieter} version of the same rubato.

\subsection{6.2 Scope and limitations}\label{scope-and-limitations}

\texttt{vega-mir} operates on \textbf{symbolic} input. It does not
transcribe audio, segment recordings, or estimate keys; that pipeline is
documented and certified upstream in
(\citeproc{ref-jalbert2026cygnus}{Jalbert-Desforges 2026}). Users who
already have transcription, segmentation, or score-aligned data (from
\texttt{music21}, \texttt{partitura}, MIDI, or the Cygnus pipeline) can
use \texttt{vega-mir} directly; users starting from audio need to plug a
transcription stage upstream.

Three of the nine metrics (multi-dimensional Gini, stationarity test,
interval-distribution fit) are presented here at the catalogue level
only and have not yet been deployed in a published study at corpus
scale. They are validated against analytic anchors and against the
Cygnus reference implementation, but their empirical behaviour on large
corpora remains to be characterised in future work.

\subsection{6.3 Reproducibility}\label{reproducibility}

The \texttt{vega-mir} release documented in this paper is permanently
archived on Zenodo with DOI
\href{https://doi.org/10.5281/zenodo.19711033}{\texttt{10.5281/zenodo.19711033}}
(v0.0.1) and is \texttt{pip\ install\ vega-mir}-able from PyPI. The
\(181\)-test suite runs on \texttt{ubuntu-latest},
\texttt{macos-latest}, and \texttt{windows-latest} against Python 3.10
through 3.13 in the continuous-integration matrix on GitHub Actions. Two
executed Jupyter notebooks (a pedagogical tour on synthetic inputs whose
analytic answers are known, and a reproduction of three flagship
findings of (\citeproc{ref-jalbert2026cygnus}{Jalbert-Desforges 2026})
on a bundled \(8\)-composer dataset) ship with the library at release
time. The analytical scripts that produce every numerical value and
every figure in this paper, plus the JSON output of each run, are
bundled at
\href{https://github.com/fredericjalbertdesforges/vega-mir/tree/main/notebooks/_arxiv_v1}{\texttt{notebooks/\_arxiv\_v1/}}:
\texttt{section3\_8\_sanity.py} (§3.8 catalogue-metric sanity checks),
\texttt{section4\_network.py} and \texttt{section4\_figure.py} (§4
network analysis and Figure 1), \texttt{section4\_diagnostics.py} (the
ablation, jackknife, and gravity-centre encoding underlying §4.3),
\texttt{section5\_rubato.py} and \texttt{section5\_figure.py} (§5 rubato
analysis and Figure 2), and \texttt{section5\_robustness.py} (§5.6
threshold sensitivity). A guided Jupyter notebook version of the §§4-5
case studies will be added in a forthcoming release alongside the next
PyPI/Zenodo bump.

\subsection{6.4 Future work}\label{future-work}

Three directions are envisaged. First, deploying the four catalogue-only
metrics (Gini, stationarity, Higuchi, intervals) in published studies at
corpus scale, closing the gap between their validated implementation and
their empirical characterisation on real data. Second, integrating with
the broader symbolic-music toolchain: \texttt{music21} and
\texttt{partitura} adapters for converting parsed scores into
\texttt{vega-mir} inputs without manual glue code. Third, extending the
network analysis from chord graphs to graphs of higher-order
\(n\)-grams, motif graphs, and rhythmic-pattern graphs, expanding the
network-theoretic surface of analysis available to symbolic-music
research.

\subsection{6.5 Conclusion}\label{conclusion}

\texttt{vega-mir} packages a documented information-theoretic
methodology as a citable, tested Python library. The two case studies
presented here show that information-theoretic descriptors beyond the
marginal level, network topology of chord transitions and spectral
colour of rubato, recover signal that scalar summaries discard. Both
descriptors are now available as one-line function calls in a public
toolkit.

\section{Acknowledgements}\label{acknowledgements}

\texttt{vega-mir} extracts the methodology already documented and
validated in the upstream Cygnus pipeline
(\citeproc{ref-jalbert2026cygnus}{Jalbert-Desforges 2026}) and packages
it as a standalone, citable library for the broader MIR community. The
author thanks the maintainers of NumPy, SciPy, NetworkX, and the Jupyter
stack on which \texttt{vega-mir} rests, and Prof.~Qiuqiang Kong for the
arXiv \texttt{cs.SD} endorsement that opened the path for both the
upstream Cygnus paper
(\citeproc{ref-jalbert2026cygnus}{Jalbert-Desforges 2026}) and the
present work.

\section*{References}\label{references}
\addcontentsline{toc}{section}{References}

\protect\phantomsection\label{refs}


\begin{CSLReferences}{1}{1}
\bibitem[\citeproctext]{ref-cancino2022partitura}
Cancino-Chacón, Carlos Eduardo, Silvan David Peter, Emmanouil
Karystinaios, Francesco Foscarin, Maarten Grachten, and Gerhard Widmer.
2022. {``Partitura: A {P}ython Package for Symbolic Music Processing.''}
\emph{Journal of Open Source Software} 7 (76): 4519.
\url{https://doi.org/10.21105/joss.04519}.

\bibitem[\citeproctext]{ref-cuthbert2010music21}
Cuthbert, Michael Scott, and Christopher Ariza. 2010. {``Music21: A
Toolkit for Computer-Aided Musicology and Symbolic Music Data.''}
\emph{Proceedings of the 11th International Society for Music
Information Retrieval Conference (ISMIR)} (Utrecht, Netherlands),
637--42.

\bibitem[\citeproctext]{ref-hagberg2008networkx}
Hagberg, Aric A., Daniel A. Schult, and Pieter J. Swart. 2008.
{``Exploring Network Structure, Dynamics, and Function Using
{N}etwork{X}.''} \emph{Proceedings of the 7th {P}ython in Science
Conference (SciPy)} (Pasadena, CA, USA), 11--15.

\bibitem[\citeproctext]{ref-harris2020numpy}
{Harris, Charles R. et al.} 2020. {``Array Programming with {NumPy}.''}
\emph{Nature} 585 (7825): 357--62.
\url{https://doi.org/10.1038/s41586-020-2649-2}.

\bibitem[\citeproctext]{ref-higuchi1988approach}
Higuchi, Tomoyuki. 1988. {``Approach to an Irregular Time Series on the
Basis of the Fractal Theory.''} \emph{Physica D: Nonlinear Phenomena} 31
(2): 277--83. \url{https://doi.org/10.1016/0167-2789(88)90081-4}.

\bibitem[\citeproctext]{ref-jalbert2026cygnus}
Jalbert-Desforges, Fred. 2026. \emph{An Audio-to-Analysis Pipeline with
Certified Transcription for Information-Theoretic Profiling of the Piano
Repertoire}. \url{https://doi.org/10.48550/arXiv.2605.06685}.

\bibitem[\citeproctext]{ref-llorens2023musif}
Llorens, Ana, Federico Simonetta, Márius Serrano, and Álvaro Torrente.
2023. {``Musif: A {P}ython Package for Symbolic Music Feature
Extraction.''} \emph{Proceedings of the Sound and Music Computing
Conference (SMC)} (Stockholm, Sweden).
\url{https://doi.org/10.48550/arXiv.2307.01120}.

\bibitem[\citeproctext]{ref-mckay2018jsymbolic}
McKay, Cory, Julie E. Cumming, and Ichiro Fujinaga. 2018. {``jSymbolic
2.2: Extracting Features from Symbolic Music for Use in Musicological
and {MIR} Research.''} \emph{Proceedings of the 19th International
Society for Music Information Retrieval Conference (ISMIR)} (Paris,
France), 348--54.

\bibitem[\citeproctext]{ref-shannon1948mathematical}
Shannon, Claude Elwood. 1948. {``A Mathematical Theory of
Communication.''} \emph{The Bell System Technical Journal} 27 (3):
379--423. \url{https://doi.org/10.1002/j.1538-7305.1948.tb01338.x}.

\bibitem[\citeproctext]{ref-virtanen2020scipy}
{Virtanen, Pauli et al.} 2020. {``{SciPy 1.0}: Fundamental Algorithms
for Scientific Computing in {P}ython.''} \emph{Nature Methods} 17:
261--72. \url{https://doi.org/10.1038/s41592-019-0686-2}.

\bibitem[\citeproctext]{ref-zipf1949human}
Zipf, George Kingsley. 1949. \emph{Human Behavior and the Principle of
Least Effort}. Addison-Wesley.

\end{CSLReferences}
\end{document}